\documentclass[fleqn,10pt]{wlscirep}
\usepackage[utf8]{inputenc}
\usepackage[T1]{fontenc}

\usepackage{amsmath}
\usepackage{hyperref}
\usepackage{graphicx}
\usepackage{bbm}
\usepackage{amssymb}
\usepackage[english]{babel}
\usepackage[normalem]{ulem}

\usepackage{soul}
\usepackage{xspace} 

\usepackage{dcolumn}
\usepackage{color}

\newcommand{\e}{\mathrm{e}} 

\newcommand{\de}{\mathrm{d}}

\title{Multipassage    Landau-Zener    tunneling   oscillations    in
  transverse/longitudinal dual dressing of atomic qubits} 

\author[1]{Alessandro Fregosi}
\author[2,1]{Carmela Marinelli}
\author[1]{Carlo Gabbanini}
\author[2]{Valerio Biancalana}
\author[3,4]{Maria Allegrini}
\author[3,1]{Ennio Arimondo}
\author[5]{Francesco Petiziol}
\author[6,7]{Sandro Wimberger}
\author[1]{Andrea Fioretti}
\author[2,*]{Giuseppe Bevilacqua}

\affil[1]{INO-CNR, Via G. Moruzzi 1, 56124 Pisa, Italy}
\affil[2]{Dipartimento  di Scienze  Fisiche,  della  Terra e  dell'Ambiente, Universit\`a degli Studi di Siena, Via Roma 56, 53100 Siena, Italy}
\affil[3]{Dipartimento di Fisica E. Fermi, Universit\`a of Pisa, Largo B. Pontecorvo 3, 56127 Pisa, Italy}
\affil[4]{NEST, Istituto Nanoscienze CNR, Piazza S. Silvestro 12, 56127 Pisa, Italy} 
\affil[5]{Technische Universität Berlin, Institut für Physik und Astronomie, Hardenbergstra\ss e 36, 10623 Berlin, Germany}
\affil[6]{Department of Mathematical, Physical and Computer Sciences, University of Parma, Parco Area delle Scienze 7/A, 43124, Parma, Italy}
\affil[7]{National Institute for Nuclear Physics (INFN), Milano Bicocca Section, Parma Group,  Parco Area delle Scienze 7/A, 43124, Parma, Italy}
\affil[*]{bevilacqua@unisi.it}

\begin{abstract}
  We  investigate the  time evolution  of a  non-resonant dressed-atom
  qubit  in  an  XZ  original  configuration. It  is  composed  of  two
  electromagnetic  fields,  one  oscillating parallel  and  the  other
  orthogonal  to   the  quantisation   magnetic  static   field.   The
  experiments   are  performed   in   rubidium   and  caesium   atomic
  magnetometers, confined  in a magneto-optical  trap and in  a vapour
  cell,  respectively.  Static  fields  in the  $\mu$T  range and  kHz
  oscillating fields  with large  Rabi frequencies are  applied.  This
  dual-dressing  configuration is  an  extension  of the  Landau-Zener
  multipassage  interferometry  in  the   presence  of  an  additional
  dressing field  controlling the tunneling process  by its amplitude
  and  phase.   Our  measurement  of the  qubit  coherence  introduces
  additional  features  to  the   transition  probability  readout  of
  standard   interferometry.    The   coherence  time   evolution   is
  characterized by  oscillations at several frequencies,  each of them
  produced  by  a  different  quantum  contribution.   Such  frequency
  description  introduces  a new  picture  of  the qubit  multipassage
  evolution.   Because the  present  low-frequency dressing  operation
  does not fall within the standard Floquet engineering paradigm based
  on  the  high-frequency expansion,  we  develop  an ad-hoc  dressing
  perturbation treatment.  Numerical simulations support the adiabatic
  and non-adiabatic qubit evolution.
\end{abstract}
\begin{document}

\flushbottom
\maketitle
%
%
\thispagestyle{empty}


\section*{Introduction}
\label{sec:introduction}

Floquet engineering is a vital technique for analyzing quantum systems
influenced by periodic electromagnetic fields, allowing the generation
of  unique  quantum  properties   across  various  domains,  including
physics,                         chemistry,                        and
engineering~\cite{Shirley_1965,Sambe_1973,GoldmanDalibard_2014,BukovPolkovnikov_2015,Eckardt_2017}. It
captures the dynamics of time-dependent Hamiltonians through effective
Hamiltonians that account for the system's key features, characterized
by two  distinct time evolutions:  a micromotion, which occurs  at the
frequency of  the periodic field,  and a slower evolution  governed by
the effective Hamiltonian.

The  experimental approach  to a  Floquet engineered  system typically
isolates the slower dynamics by  integrating the faster, less relevant
one, often eliminating the need for  a detailed integration due to the
small amplitude  of the  micromotion. This capability  is advantageous
for  quantum simulation  tasks, such  as those  involving cold  atomic
gases
~\cite{AidelsburgerBloch2013,MeinertNaegerl2016,GeierWeidemueller2021,WeitenbergSimonet2021,SchollBrowaeys2022,YinChang2022,ZhangPan2023} 
and                             solid                            state
ones~\cite{SalatheWallraff2015,MahmoodGedik2016,PengCappellaro2021}. There,
the intricacies  of the micromotion can  be neglected to focus  on the
emergent phenomena captured by the effective Hamiltonian.

We present  here experimental  and theoretical  studies on  the strong
periodic bi-modal nonresonant drive of a two-level atomic qubit, which
fits within  the broader  context of  two-level systems  influenced by
virtual/real
photons~\cite{AllegriniArimondo_2024,PassanteRizzuto_2025}.       This
concept,     initially     proposed     by     Cohen-Tannoudji     and
Haroche~\cite{HarocheCohen_70_1,HarocheCohen_70_2},   focuses  on   an
energy  splitting  that  is  small relative  to  the  driving  field's
frequency.  In  the  Floquet   framework,  the  effective  Hamiltonian
accounts  for an  energy  splitting modulated  by  the dressing  field
amplitude.                                                    Previous
work~\cite{BevilacquaArimondo_2020,BevilacquaArimondo_2022}  exploring
dual-dressing   with   two   electromagnetic   fields   of   different
polarizations has  provided insight  into how these  dual interactions
can  yield new  dynamical behaviours  and effective  Hamiltonians that
significantly modify the system's  energy landscape.  Several original
features are introduced into the present dual dressing study.

In our configuration,  an atomic qubit with  energy splitting produced
by  a static  magnetic field  is  dressed by  oscillating fields  with
frequencies  lower than  the qubit's  Larmor precession  frequency and
large  amplitudes.  Here,  the   dynamics  unfolds  a  fast  timescale
diverging from  the standard  Floquet treatment that,  unlike previous
studies    indicating    adiabatic    behaviour   in    slow    driving
regimes~\cite{BukovPolkovnikov_2015},    allows    us    to    explore
non-adiabatic features.  Our focus  on quantum control involves direct
measurement of  the qubit  time evolution,  revealing a  distinct spin
evolution that deviates from traditional micromotion patterns.

The XZ dual-dressing  setup utilizes the interaction  of atomic qubits
with two off-resonant magnetic  fields, namely an $x$-axis oscillating
field  and  a  $z$-axis  oscillating one  periodically  adjusting  the
magnetic  energy gap.   XZ dual  dressing  of a  two-level system  was
proposed in Refs.~\citeonline{PetiziolWimberger_2018,PetiziolWimberger_2024}
within  the  context  of   nonadiabatic  dynamics  acceleration.  This
configuration implements  a periodic  multi-passage scheme akin  to the
Landau-Zener-St\"{u}ckelberg-Majorana   (LZSM)    interferometer,   as
reviewed in Ref.~\citeonline{ShevchenkoNori_10,IvakhnenkoNori_2023},
examined                 in                solid                 state
experiments~\cite{OliverOrlando_05,SillanpaaHakonen_06,ZhouDu_14}  and
planned in  optical lattice clocks~\cite{LiuLi_21}.   Our experimental
approach enhances LZSM studies by detecting the transverse oscillating
spin  component  rather  than the  conventional  probability  transfer
between  eigenstates.  The  relative   phase  of  the  driving  fields
regulating  the  periodicity of  the  multi-passage  tunneling and  the
continuous monitoring  represent additional LZSM handles.   The strong
nonlinear   response  of   the  spin   dynamics  produces   high-order
interference oscillations with their collapse and revival at different
time periods.  The  direct detection of the qubit  full wave function,
including its  phase, is  important nowadays  for quantum  control and
quantum information, as pointed out in Ref.~\citeonline{KofmanNori_2023}.
 
Operating at high electromagnetic field strengths, our system deviates
from the rotating-wave approximation, used typically in LZSM analyses.
For a  two level driven  system the electromagnetic field  strength is
usually  measured by  the  ratio  between the  Rabi  frequency of  the
electromagnetic drive and the Larmor frequency. While experiments with
a   pulsed   laser   excitation  hardly   account   phase   relations,
continuous-wave   experiments  mitigate   these  concerns.   This  was
demonstrated                          in                         prior
works~\cite{FuchsAwschalom_2009,TuorilaHakonen_2010,ScheuerJelezko_2014,AvinadavGershoni_2014,DengLupascu_2015}
exploring Rabi frequencies up to four times the Larmor frequency, with
the     highest    values     reached    inside     a    cavity     as
in Ref.~\citeonline{LangfordDiCarlo_2017}.  In this work we employ Rabi
frequencies  up  to seven  times  larger  than the  Larmor  frequency,
significantly enhancing quantum control capabilities. 

The experiments,  exploiting the setups of  the previous investigation
of   the   XY   dual   dressing    on   both   cesium   and   rubidium
atoms,~\cite{BevilacquaArimondo_2020,FregosiFioretti_2023,BevilacquaFregosi_2025}
are briefly described in
Section ``The qubit system and its detection''.  Here, we also present
the basic Hamiltonian and some  typical evolutions of the system.  The
full theoretical description, presented in
Section ``Theoretical treatment'' is based  on several tools.  For the
low magnetic fields  of our explorations, the atoms  are equivalent to
an assembly  of one-half  spins.  Owing  to the  reversed role  of our
dressing/Larmor  frequencies,  the   standard  Floquet  high-frequency
expansion  (HFE) approach  cannot be  applied.  Therefore,  we present
theoretical descriptions based on i)  a full adiabatic evolution valid
for weak dressing parameters,  ii) a quasi-adiabatic evolution leading
to an  analytical solution for the  spin's time evolution, and  iii) a
modified Floquet treatment.  Numerical simulations describe the strong
field dynamics.  The connection  with the St\"{u}ckelberg oscillations
and other  LZSM features is  emphasized.  The experiments  monitor the
time dependence of  the atomic spin evolution and access  the phase of
its   wavefunction.    Several    recorded   atomic   evolutions   are
presented.
Section  ``LZSM data  analysis'' discusses  the interpretation  of our
results and  the theoretical comparison to  the experimental findings.
The ``Conclusion'' Section completes our work.

\begin{figure}
\centering
\includegraphics[width=0.6\textwidth]{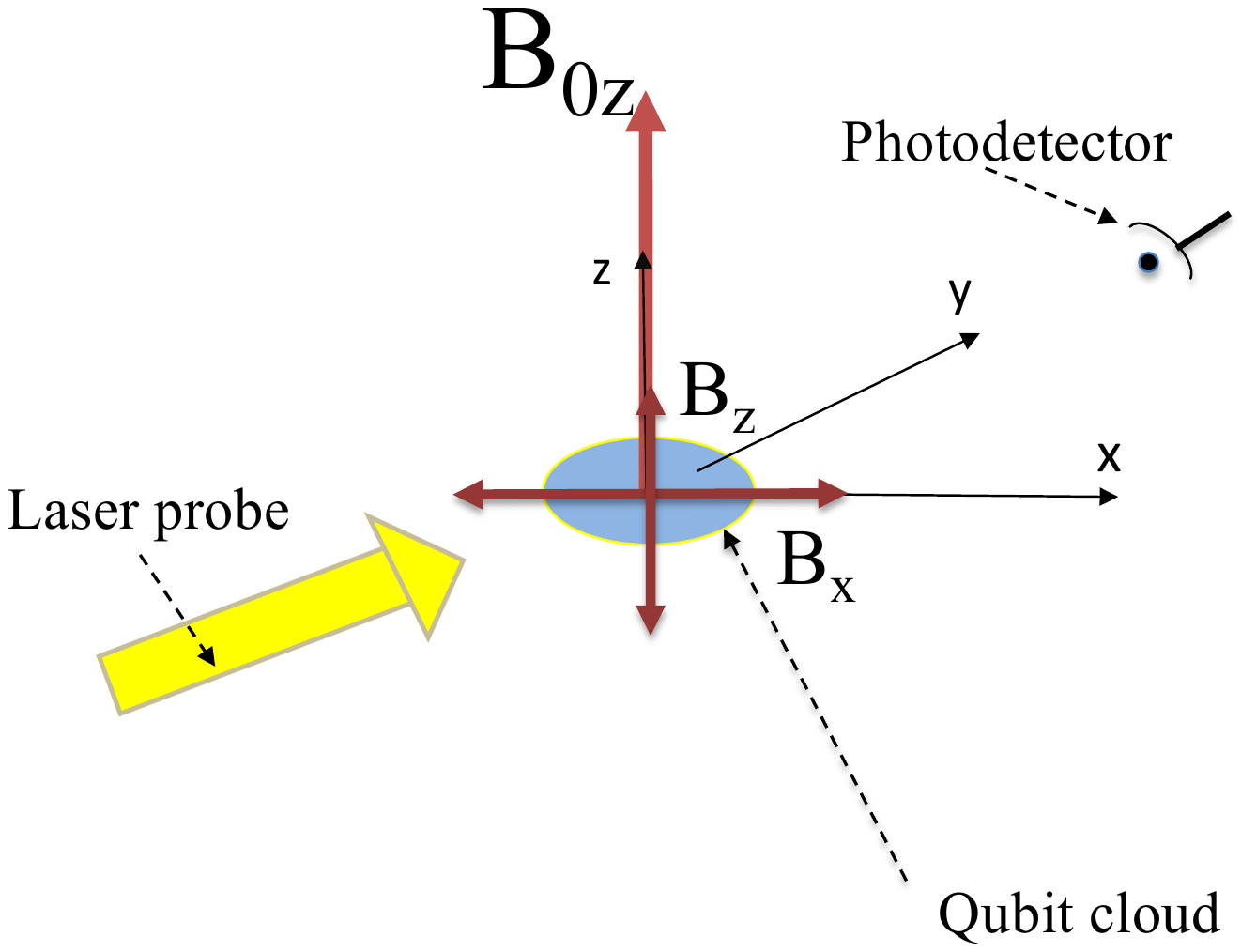}
\caption{Schematic  of a qubit dressed
      by  the $B_x$  and $B_z$  oscillating fields,  generated by  the
      radiofrequency coils  and in the  presence of a  $B_{0z}$ static
      field.   In  this  figure   the  $\langle  \sigma_y(t)  \rangle$
      expectation value is monitored by the polarization rotation of a
      probe beam propagating along the $y$ axis. } 
\label{fig:setup}
\end{figure}

\section*{The qubit system and its detection}
\label{sec:experiment}
\subsection*{Hamiltonian}

The time evolution of an  assembly of atomic non-interacting qubits in
a dual dressing configuration  is explored as in Fig.~\ref{fig:setup},
i.e., in the presence of a  static magnetic field $B_{0z}$ parallel to
the $z$-axis  and two non-resonant electromagnetic  fields oscillating
at $\omega$  angular frequency. For  the XZ dressing these  fields are
oriented along the $x$ and $z$  axes, with $B_i$ the maximum value for
$i=(x,z)$. Similar results are obtained  in the equivalent YZ dressing
configuration. At the
applied magnetic fields 
the atomic structure  of rubidium/caesium atoms can be  described by a
collection of degenerate two-level systems, representing the qubits.

The  qubit-field  coupling  is  determined  by  the  $\gamma=g\mu_B  $
constant with  $g$ an  effective Land\'e factor  and $\mu_B$  the Bohr
magneton,  assuming  $\hbar=1$.   The  spin-static  field  interaction
corresponds to  the $\omega_{0z}=\gamma B_{0z}$ Larmor  frequency. The
coupling  with  the  XZ  dressing  fields is  described  by  the  Rabi
frequency amplitudes $\Omega_x=\gamma  B_x$ and $\Omega_z=\gamma B_z$,
respectively.  Introducing  $\Phi_{0z}$, the phase  difference between
$x$ and $z$ dressings, the XZ Hamiltonian is written as
\begin{equation}
\label{eq;Hamiltonian}
H(t)= \frac{1}{2} \vec{h}(t) \cdot \vec{\sigma},
\end{equation}
with the effective $\vec{h}(t)$ field given by
\begin{equation}
  \label{eq:def:h}
  \vec{h}(t) = 
  \begin{pmatrix}
   \Omega_x \cos(\omega t)\\
    0\\
    \omega_{0z}+\Omega_{z}  \cos(\omega t+\Phi_{0z})
  \end{pmatrix}.
  \end{equation}

Owing  to   negligible  decoherence  processes  this   Hamiltonian  is
complete. Starting  from the initial  $\langle\sigma_x(t=0) \rangle=1$
qubit  coherence,  the  time  evolution of  the  $\langle  \sigma_x(t)
\rangle$ coherence,  or of the equivalent  one along the $y$  axis, is
monitored  in  the  experiment.  The theory  comparison  requires  the
control  of  the  following   parameters:  the  oscillation  frequency
$\omega$,  the  bare  Larmor   angular  frequency  $\omega_{0z}$,  the
$\Omega_x$,  (transverse)  and   the  $\Omega_z$  (longitudinal)  Rabi
angular frequencies, all expressed in  kHz units in the following, and
the     $\Phi_{0z}$    phase.     As    for     our    $\omega     \ll
\omega_{0z},\Omega_x,\Omega_z$  operational  parameters  the  standard
dressed atom HFE approach used in Floquet engineering is not suitable,
we  rely  on various  theoretical  treatments:  1) an  adiabatic  one,
improved by non-adiabatic corrections, 2) a perturbative approach in a
suitable rotating  frame owing  to the large  $\Omega_z/\omega$ ratio,
and finally 3) numerical simulations.

We notice  that the  above XZ  bichromatic driving  Hamiltonian, which
reduces  to the  LZSM one  for  $\Omega_z=0$, represents  a very  rich
configuration  to  explore. A  LZSM  $\Omega_z=0$  scheme with  Z-axis
bichromatic driving was explored in Ref.~\citeonline{ForsterKohler_15}.  
 
\subsection*{Experimental setup and results}
\label{subsec:setup}
We perform  our observations in  two experimental setups: a  first one
where  an $^{85}$Rb  atomic sample  is laser-cooled  and trapped  in a
magneto-optical  trap to  a few  tens of  micro-Kelvins, occupying  the
single  $F_g=3$  hyperfine  state~\cite{FregosiFioretti_2020},  and  a
second apparatus  that employs $^{133}$Cs  atoms contained in  a vapor
cell      and     prepared      in      the     $F_g=4$      hyperfine
state~\cite{Bevilacqua_apl_2019}.  In the  Rb  case,  the cold  atomic
sample is monitored  during its 10 ms free-fall and  exhibits a damped
dynamics with a  characteristic time of about 4~ms. In  the Cs case, a
vapor sample in 23~Torr buffer gas  contained in a centimetric cell is
interrogated  after  being  optically  laser pumped,  and  the  signal
damping  lasts about  10~ms.  Both  experiments work  in cycles:  in a
first phase atoms are prepared spin-polarized along the x-axis using a
circularly polarized pump laser in  the presence of a uniform $B_{0z}$
static magnetic field.  
At  the  conclusion  of  the polarization  phase,  two  radiofrequency
fields, linearly  polarized and operating  in the 1-10 kHz  range with
amplitudes varying from 0 to 50 $\mu$T, are applied to the atoms along
the  ({\it x,  z})  or ({\it  y,  z}) directions.  The  time scale  of
dressing  evolution refers  to the  zero initial  phase choice  of the
effective field, as described in Eq.~\eqref{eq:def:h}.

The expectation value of  $\langle \sigma_y(t) \rangle$ for Rb
and $\langle \sigma_x(t) \rangle$ for Cs is monitored by examining the
polarization  rotation (Faraday  effect) of  a probe  beam propagating
along  the {\it  y}  or  {\it x}-axis,  respectively,  as depicted  in
Fig.~\ref{fig:setup}.   

The  rotation  angle,   from  the  initial  direction   to  the  final
orientation,  is analyzed  with a  balanced polarimeter,  whose output
enables  a  precise  detection  of  the amplitude  and  phase  of  the
magnetization component along  the monitored axis.  In  both cases the
measured Faraday rotation signal is renormalized to compensate for the
signal  decay. The  time resolution  is limited  by the  sampling rate
which  is  250~kHz  and  125~kHz  in  the Rb  and  in  the  Cs  cases,
respectively.  The Rb  experiment  uses crossed  laser  beams for  the
pumping  and  probe purposes,  both  tuned  to  the  $D2$ line  of  Rb
(780~nm), while the  Cs one uses two co-propagating  laser beams tuned
to $D$1  (pump, 895~nm) and $D$2  (probe, 852~nm) lines of  Cs. In the
cesium case the pump component is detuned several GHz out of resonance
during  the   measurement  and   finally  filtered  away   before  the
polarimeter while it is just cut off in the rubidium experiment.

\begin{figure}[ht!!]
  \centering
  \includegraphics [angle=0, width=0.33\textwidth,height=4cm]{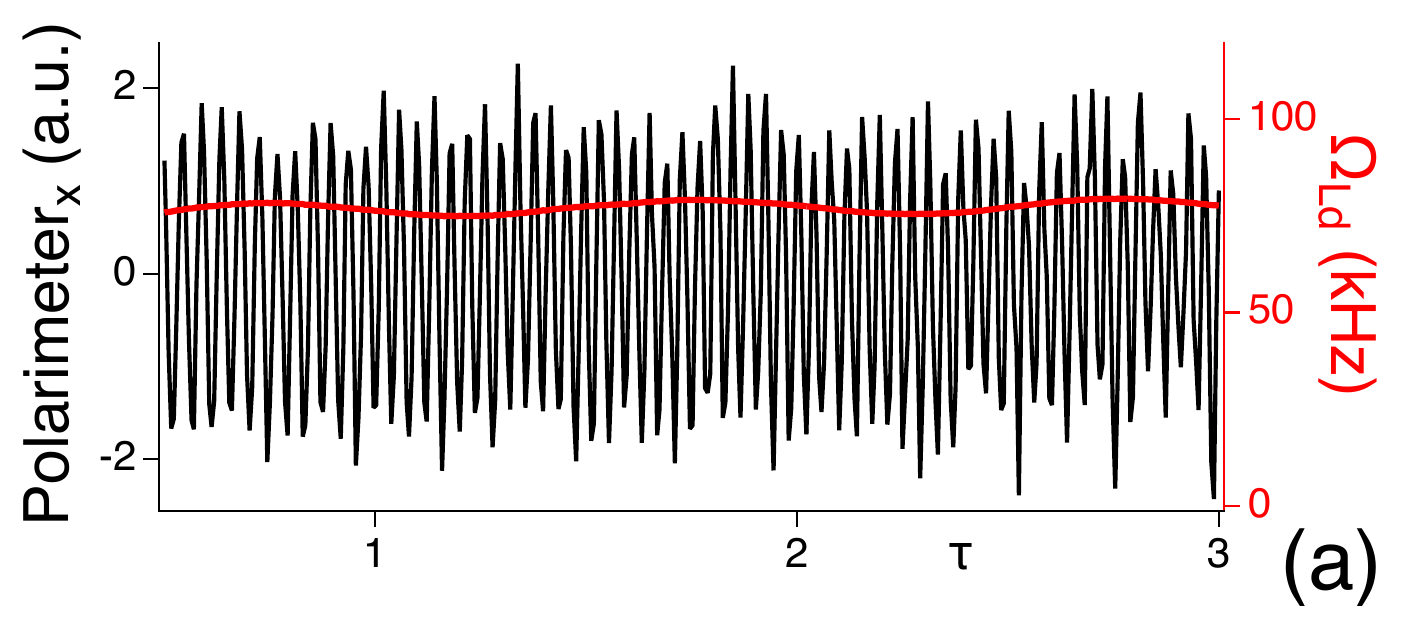}  
  \includegraphics [angle=0, width=0.33\textwidth,height=4cm]{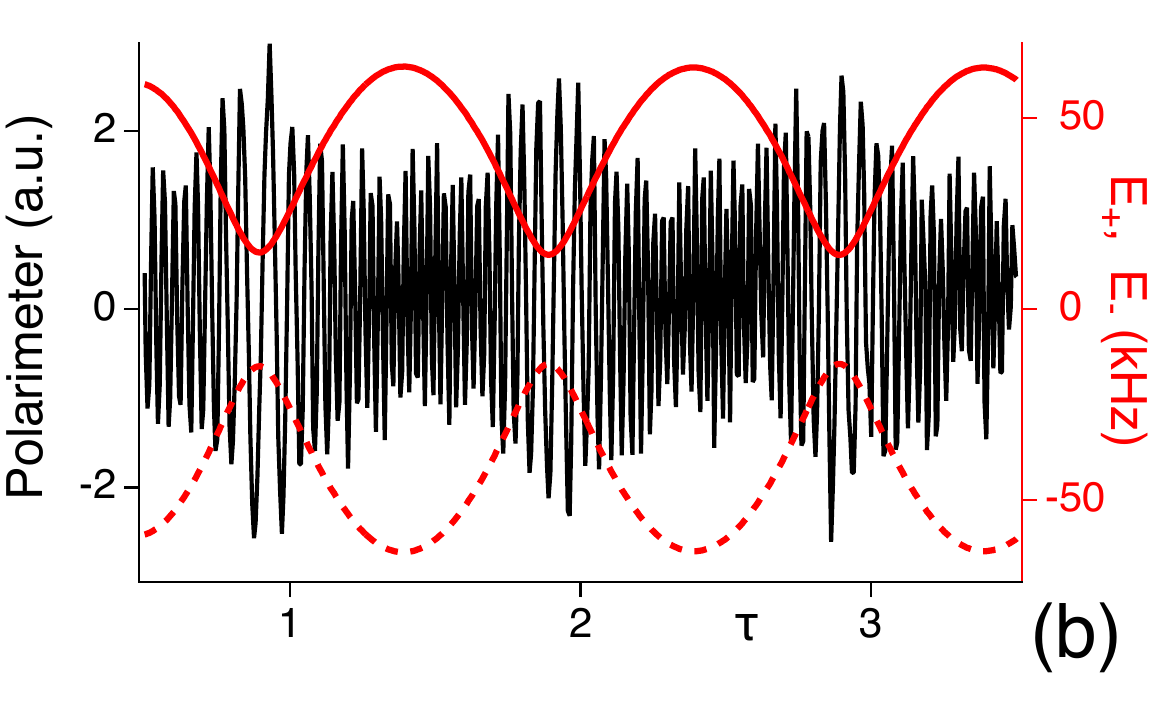}  
  \includegraphics [angle=0, width=0.33\textwidth,height=4cm]{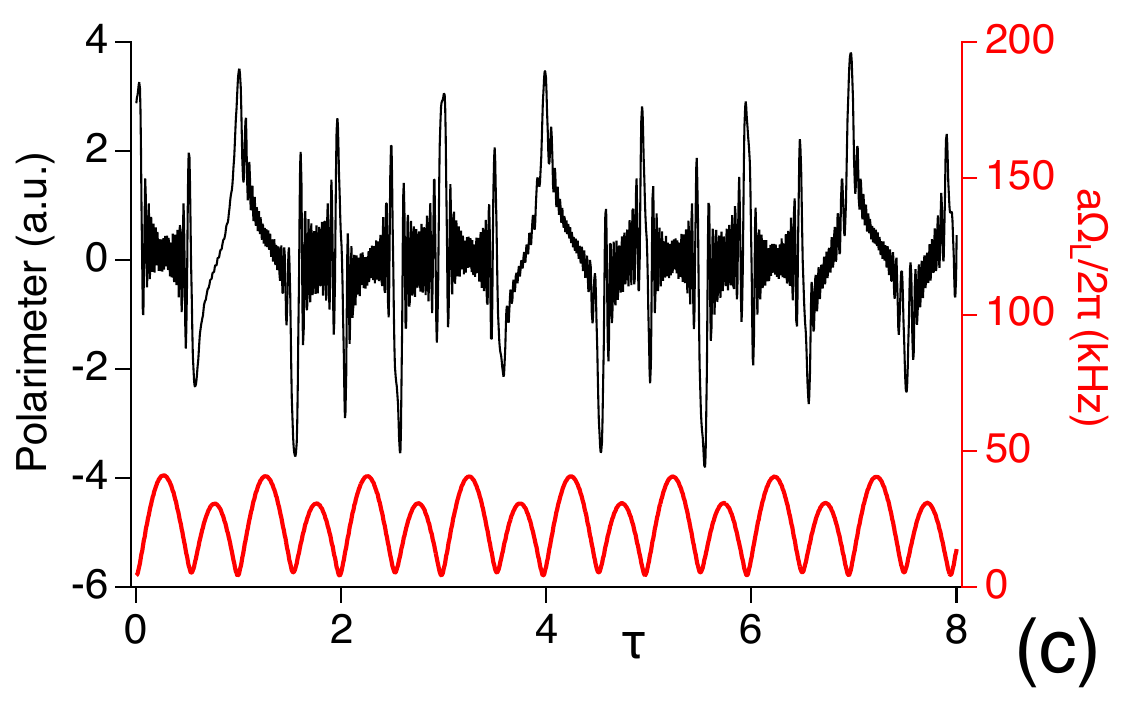}  
  \caption{(Color online)  Time evolution  of the atomic  spin.  Black
    lines  (left axis)  report experimental  polarimeter signals,  red
    lines    (right   axis)    the   theoretical    $\Omega_{Ld}$   of
    Eq.~\eqref{eq:XZ:htheta:def} derived  from the applied  static and
    oscillating  field  values.  On  the horizontal  axis  the  $\tau$
    reduced time.  Parameters ($\omega,\omega_{0z},\Omega_x,\Omega_z$)
    all in kHz  and $\Phi_{0z}$. In (a) Rb atoms,  (3.0, 77.645, 2.06,
    2.0) and $\Phi_{0z}=0$, a linearly polarized oscillating field; in
    (b) Rb atoms, (3.0, 77.645, 53.1, 146.9) and $\Phi_{0z}=\pi/2$; in
    (c) Cs atoms, (1.028,  4.42, 6.27, 31.79) and $\Phi_{0z}=0.70\pi$.
    For the (c) parameters the large value of the Z dressing amplitude
    leads to  two maxima  and minima  within each  time period  of the
    effective field.  Note the different time periodicity.}
  \label{fig:XZTime}
\end{figure}

\begin{figure}[ht!!]
  \centering
  \includegraphics   [angle=0,   width=  0.45\textwidth,   height=8cm]
  {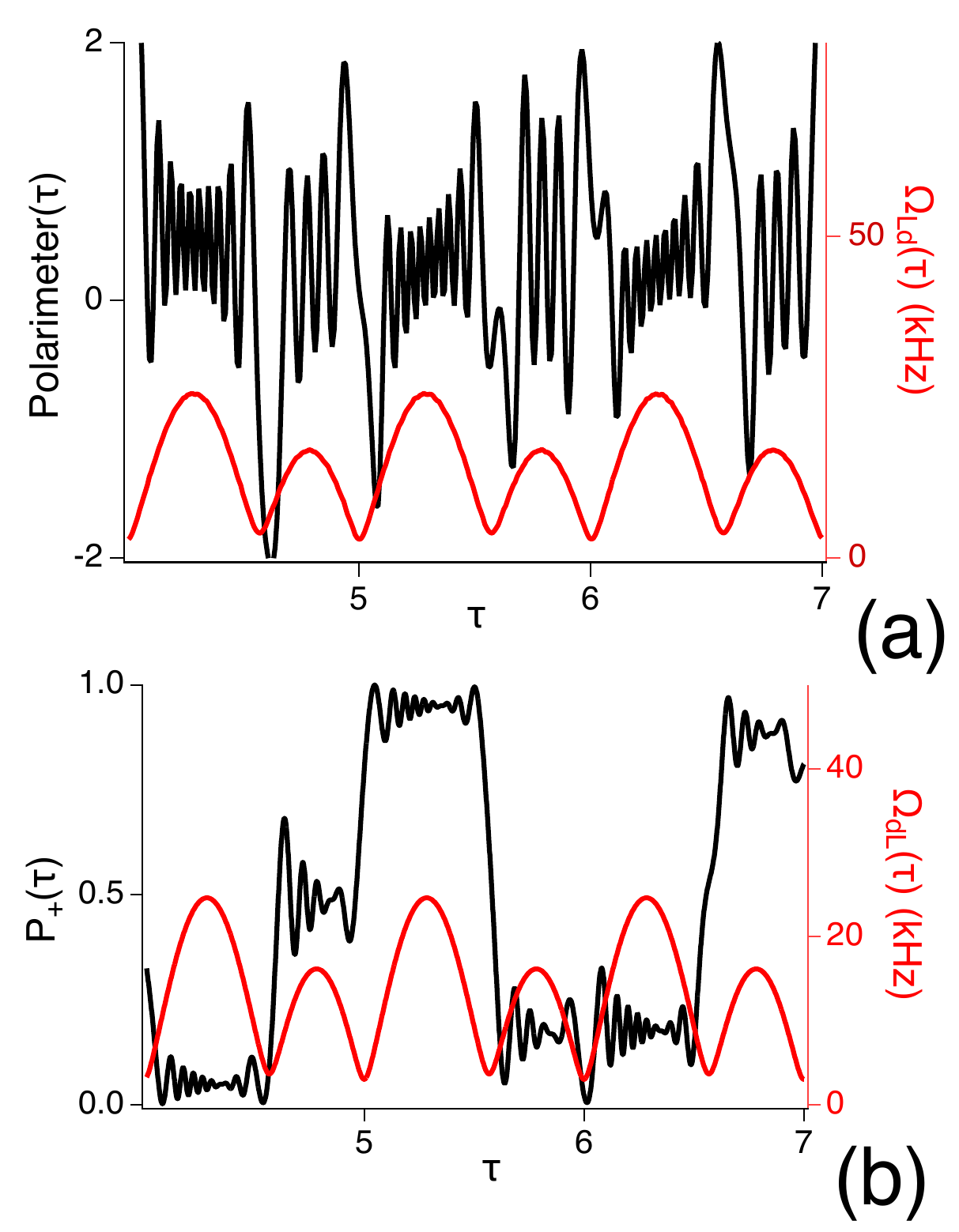}
  \includegraphics   [angle=0,   width=  0.45\textwidth,   height=8cm]
  {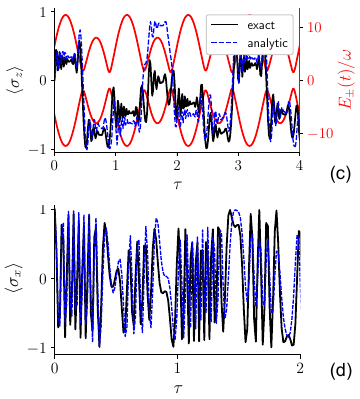}
  \caption{(Color online) Left column:  in (a) black line experimental
    polarimeter signal and in (b)  $P_+$ probability occupation of the
    $|+\rangle$  eigenstate derived  from  numerical  solution of  the
    Schr\"odinger equation  as function of the  $\tau=\omega t/(2\pi)$
    reduced  time.   Red   lines:  in  the  time   dependence  of  the
    $\Omega_{Ld}$ Larmor dressed  frequency.  Right column: comparison
    between  the exact  numerical  time evolution  and the  analytical
    estimate  [see  the  main  text   and  Appendix  ''Model  for  the
    low-frequency nonadiabatic regime''] based on the modified Floquet
    approach.    Cs    parameters   in   kHz:   $\omega    =   1.028$,
    $\omega_{0z}=  4.42$, $\Omega_x=3.85$,  and $\Omega_z=20.93$,  and
    $\Phi_{0z}= 1.63\pi$  corresponding to  a nonadiabatic  regime.  Notice
    that  for these  parameters  the  large value  of  the Z  dressing
    amplitude leads to  two maxima and minima within  each time period
    of the effective field. }
 \label{fig:34}
 \end{figure}

Some  measured time  dependencies  of the  atomic  spin evolution  are
illustrated  in  Fig.~\ref{fig:XZTime},  with  Rabi  frequency  values
increasing from
left to right.  The balanced polarimeter output is plotted against the
reduced   time  $\tau=t/T$,   where  $T=2\pi/\omega$   represents  the
radiofrequency field  period, and $\tau=0$ corresponds  to the initial
preparation state  $\langle \sigma_x \rangle  = 1$. The plots  of this
Figure  offer   a  synthetic   overview  for  all   recorded  temporal
structures.  The  spin  component  aligned  with  the  detection  axis
oscillates  quasi-periodically between  positive and  negative values,
produced  by spin  precession in  the $(x,y)$  horizontal plane.   The
recorded  signals exhibit  three distinct  periodic or  quasi-periodic
patterns. The  first precise periodicity matches  the dressing fields'
period $T$, corresponding to $\Delta  \tau$ integer values.  The other
two patterns are shorter period structures, appearing quasi-regular in
(a) and chirped in frequency in  (b), that reflect the large amplitude
oscillations  of   the  qubit   coherence.  The  amplitude   of  those
oscillations is time modulated with  the $T$ period.  At high dressing
frequencies $\Omega_x$ or $\Omega_z$,  the oscillation pattern repeats
with a substantial $\Delta\tau$ period,  3.00(1) for the parameters in
plot   (c).    Similar   experimental   results   are   presented   in
Fig~\ref{fig:34}(a).

For the parameters in Fig. 2(c),  the ratio between Rabi frequency and
atomic splitting  is 7.4. Experiments  with ratios  up to 10  showed a
comparably  long  periodicity.   For  comparison we  notice  that  the
continuous-wave,          strong-drive          experiment          in
Ref.~\citeonline{AvinadavGershoni_2014} applied Rabi frequencies up to
four times the Larmor frequency.

\section*{Theoretical treatment}
\label{sec:theory}

Our  theoretical  analysis of  the  qubit  response follows  different
action lines.   The numerical  solution of the  Schr\"odinger equation
for  the  $|\psi\rangle$  wavefunction  represents  the  most  precise
approach.  That solution  handles also  the  more general  case of  an
initial $t=0$  qubit preparation  not matching the  zero phase  of the
$x$-axis     driving    as     in    the     effective    field     of
Eq.~\eqref{eq:def:h}.   Numerical  solutions   of  the   Schr\"odinger
equation  are  presented   in  Figs.~\ref{fig:34}(b)-(d),   and
\ref{fig:RabiFreq}.  The  numerical  solutions, representing  the  key
element for the analysis of the  experimental data, point out the high
sensitivity of  the qubit  time evolution  to dressing  parameters and
initial  conditions.   Both  elements   limit  the  accuracy   of  the
experiment-theory comparison.

Within an adiabatic  approach, the qubit time  evolution is determined
by  the  time-periodic  $(E_+(t),E_-(t)   )$  eigenvalues  of  the  XZ
Hamiltonian  of  Eq.~\eqref{eq;Hamiltonian}.   These  eigenvalues  are
derived from the $h$ modulus as follows: 
\begin{subequations}
\label{eq:XZ:htheta:def}
\begin{align}
  \Omega_{Ld}(t)&=h=\sqrt{\Omega^2_x\cos(\omega t)^2+\left[\omega_{0z}+\Omega_z\cos(\omega t + \Phi_{0z})\right]^2}, \\
  E_+(t)&=-E_-(t)= \frac{\Omega_{Ld}}{2}.
\end{align}
\end{subequations}
with the  time-dependent $\Omega_{Ld}$  dressed atom  Larmor frequency
corresponding   to  the   dressed-atom   energy  gap   at  time   $t$.
$\Omega_{Ld}$ is equivalent to the  Rabi oscillation frequency of LZSM
theory/experiment~\cite{AshhabNori_07,ZhouDu_14}. The  time dependence
of  the  $\Omega_{Ld}$ Larmor  dressed  frequency  are represented  by
continuous   red   lines   in   several   plots.   
In Fig.~\ref{fig:XZTime}(a) where the dressing amplitudes are very low
$\Omega_{Ld}$  is  almost  constant, dominated  by  the  $\omega_{0z}$
static   contribution.  In   Fig.~\ref{fig:XZTime}(b),  at   increased
dressing  amplitude,   the  modulated  component  acquires   a  larger
weight. Here, the time evolution of the eigenenergies is characterized
by  a sequence  of avoided  crossings  with $T$  periodicity. For  the
Fig.~\ref{fig:XZTime}(c) plot parameters, corresponding to even larger
modulation  amplitude,  the  dual  dressing  operation  leads  to  two
separated avoided crossings within a single $T$ period.

An  adiabatic perturbation  treatment is  valid for  a slowly-changing
time-dependent        Hamiltonian,        as        recalled        in
Appendix ``Adiabatic perturbation theory''.
It is applied in
Appendix ``XZ rotating Hamiltonian''
to  a   simplified  XZ  excitation  scheme,
denoted    as    rotating    XZ,   where    $\Omega_x=\Omega_z$    and
$\Phi=\pi/2$. This analysis shows that  the evolution of the monitored
spin components depends on two parameters: the $\theta(t)$ orientation
angle  of  the  $\vec{h}$  vector  and  the  LZSM  interference  phase
$\phi(t)$ given by 
 \begin{eqnarray}
\label{eq:XZ:Eigen:def}
  \theta(t) &=& \arctan\left( \frac{h_x}{h_z} \right) =
           \arctan
           \left( \frac{\Omega_x\cos(\omega t )}{\omega_{0z}+\Omega_z \cos(\omega t +\Phi_{0z})} \right), \nonumber\\
  \phi(t) &=&   \int_0^{t} \Omega_{Ld}(\tau) d\tau. 
\end{eqnarray}
A similar geometric approach based on a rotation around an axis in the
$xy$ plane followed by a rotation around the $z$-axis was explored for
a standard LZSM evolution in Ref.~\citeonline{ShevchenkoNori_10}.

As  in  Eq.~\eqref{eq:XZ:si:transf:vm}  the  time  dependence  of  the
$\langle\sigma_x(t)\rangle,\langle\sigma_y(t)\rangle$ spin mean values
is given by 
\begin{eqnarray}
\label{eq:XZ:sigma:transf:vm}
  \langle \sigma_x(t) \rangle & \propto & \cos(\phi),\nonumber \\ 
  \langle \sigma_y(t) \rangle & \propto & \sin(\phi).
\end{eqnarray}

These expressions, verified numerically,  evidence the key role played
by  the  $\Omega_{Ld}$ frequency  into  the  time-dependent spin  mean
values.   The qubit  oscillation  at the  $\Omega_{Ld}$ Larmor  angular
frequency   of   Eq.~\eqref{eq:XZ:htheta:def}  produces   the   $\phi$
accumulated phase.  The $\theta(t)$  orientation angle  determines the
amplitude of  the qubit oscillations monitored  along the experimental
detection axis.   The observed  complex qubit oscillating  response is
based on  the $\Omega_{Ld}$ time dependent  fast oscillations combined
with a slower scale variations of the amplitude variations. Additional
interference  oscillations  are  produced by  the  multi-passage  LZSM
avoided crossings, as shown in the following Section. \\
\indent To describe the nonadiabatic low-frequency  regime for our $\omega \ll
\omega_{0z}$  operation,  the standard  HFE  Floquet  approach is  not
appropriate. Hence, a perturbative  dressed atom approach is developed
here.  The  spirit  of  this  approach is  to  derive  a  perturbative
expansion  in  the limit  of  large  $\Omega_z^{-1}$. As  detailed  in
Appendix ``Model for the low-frequency nonadiabatic regime'',
this is  achieved by  representing the
problem  in   a  time-dependent  rotating  frame   where  the  Fourier
components  of  the Hamiltonian  are  proportional  to $\sim  \Omega_x
J_n(\Omega_z/\omega)$, where $J_n(\cdot)$ are  Bessel functions of the
first kind.  The small  value of  the latter for  large values  of the
argument  $\Omega_z/\omega$, as  compared to  the value  of $\Omega_x$
considered,   justifies  an   expansion  of   the  effective   Floquet
Hamiltonian  $H_{\rm  eff}$  and  the  Floquet  kick  operator  $K(t)$
(Appendix ``Model for the low-frequency nonadiabatic regime'').
This provides an approximation of the
evolution operator in the form of Floquet's theorem 
\begin{equation}
U(t, t_0) = e^{-i K(t)} e^{-i H_{\rm eff}(t-t_0)} e^{iK(t_0)}.
\end{equation}

As shown in Fig.~\ref{fig:34} (c)-(d), this  approach succeeds in capturing the
key features  of the  exact dynamics  with a  second-order description
also in the low-frequency, but yet nonadiabatic, regime considered.

\section*{LZSM data analysis}
\label{sec:analysis}

In a double  Landau-Zener tunneling process, with  an avoided crossing
region  passed twice  at the  same speed,  the excitation  probability
becomes  an  oscillating  function   characterized  by  the  so-called
St\"uckelberg oscillations. In a multi-passage Landau-Zener passage the
quantum-mechanical interference  for the amplitudes of  quantum states
sequentially  mixed   at  separated  crossings  leads   to  additional
structures  in  the  qubit  time-evolution.  In  the  absence  of  the
$\Omega_z$
dressing   the  Hamiltonian   of  Eq.~\eqref{eq;Hamiltonian}
reduces to  the LZSM one~\cite{ShevchenkoNori_10,IvakhnenkoNori_2023},
with the
$\Omega_x$
driving producing the  avoided-crossing sequence.
The
$\Omega_z$ coupling presence modifies the LZSM tunneling process owing
to  the time  dependent  orientation angle  of  the periodic  magnetic
structures in the qubit time-evolution.

The LZSM tunneling treatment is characterized by different operation
regimes, denoted as adiabatic  and nonadiabatic ones. For $\Omega_z=0$
the LZSM derivation of those regimes requires $\omega\Omega_x$ smaller
or              larger              than              $\omega_{0z}^2$,
respectively~\cite{ShevchenkoNori_10}.   For  $\Omega_z   \ne  0$,   a
modification of that treatment leads to 
 \begin{equation}
 \omega\sqrt{\Omega_x^2+\Omega_z^2} \lessgtr  \omega_{0z}^2.
 \end{equation}
for adiabatic and nonadiabatic evolutions, respectively.

The nonadiabatic  tunneling enhances the interference  oscillations in
the  occupation  probabilities  of the  $|\pm\rangle$  states.   Those
oscillations   appear   in   the   $P_+$   numerical   simulation   of
Figs.~\ref{fig:34}  (b)  and  (c) for  nonadiabatic  parameters.   The
overall  occupation time  dependence agrees  with previous  LZSM work.
The   probability  steps   of  our   simulations  correspond   to  the
Landau-Zener  crossing processes  at  the minima  of  the energy  gap,
visible  in  the  red  lines depicting  the  adiabatic  potentials  in
Fig.~\ref{fig:XZTime}.   Those  jumps  are followed  by  St\"uckelberg
oscillations.  These oscillations with a quasi regular frequency match
the     high    frequency     ones     appearing     in    all     the
$\langle  \sigma_x  \rangle$  plots  of  Fig.~\ref{fig:XZTime}.   Such
matching  of  the  St\"uckelberg   oscillations  applies  to  all  our
theoretical plots. These nearly periodic oscillations do not appear in
previous  St\"uckelberg   results,  for  instance  from   atom  optics
interferometry                                                      of
Refs.~\citeonline{KingerWeitz_10,ZenesiniArimondo_10,RamanGupta_24}.
This  quasi-periodicity  arises  because  the  $\omega$  rate  of  the
Landau-Zener  tunneling   is  small  compared  to   the  St\"uckelberg
oscillation frequency.
\begin{figure}[ht!]
  \centering
 \includegraphics [angle=0, width= 0.9\textwidth,height=8cm]{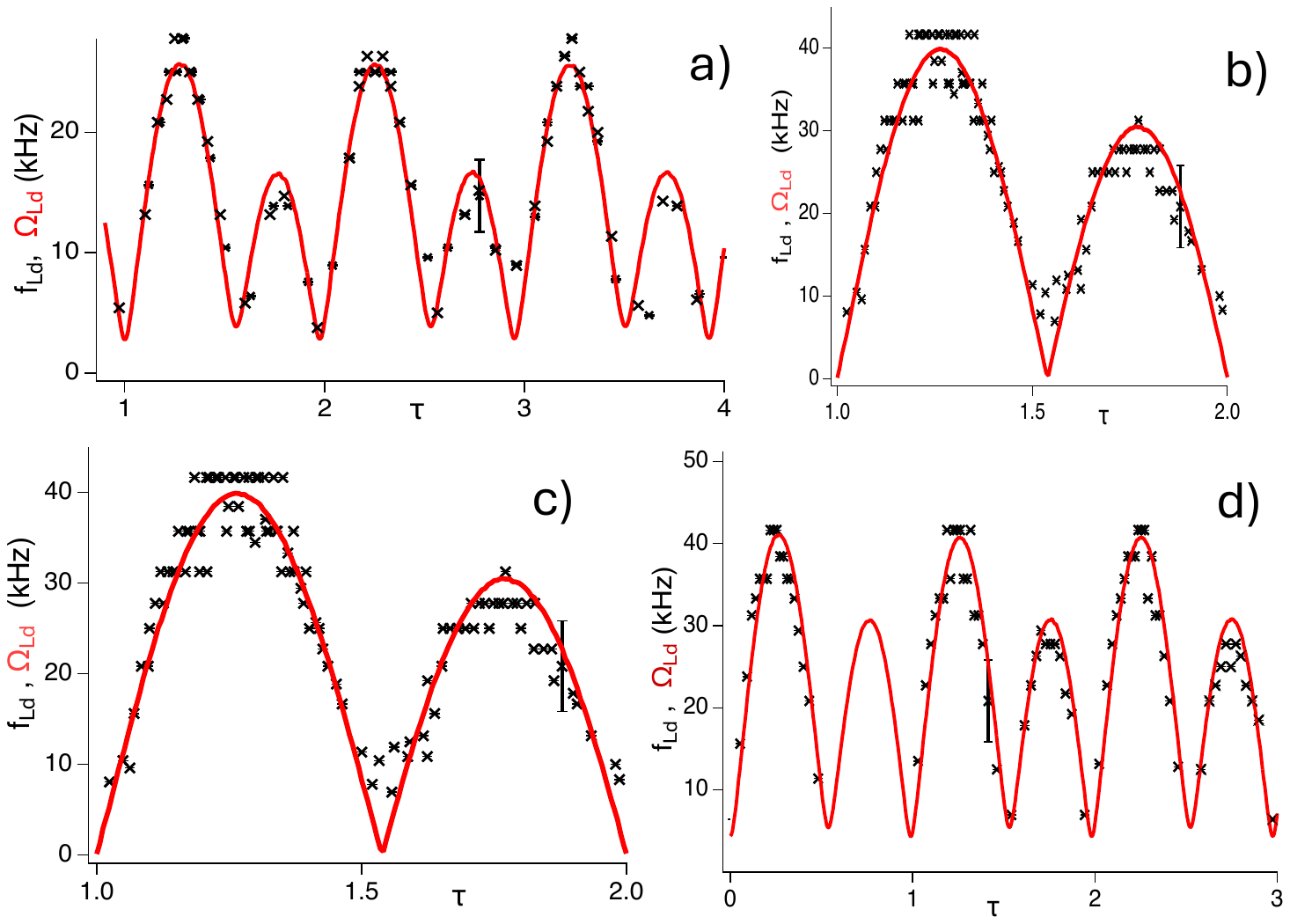}
\caption{(Color online)  Measured $f_{Ld}$ and  predicted $\Omega_{Ld}$
  values   for   time   dependent   adiabatic   frequency   versus
  time. Frequencies in kHz and times $\tau$ in reduced units. $f_{Ld}$
  frequencies (black  squares) are  derived from the  polarimeter zero
  values as  in Eq.~\eqref{eq:afL}.  The typical error  bars, reported
  for clarity only for one/two points  in each plot, are determined by
  the sampling rate,  leading to larger errors at  closely spaced time
  intervals  and higher  $f_{Ld}$ values.  In  (b) and  (c) plots  the
  $f_{Ld}$  high  value  data  are discrete  because  of  the  limited
  sampling rate.   Red lines  $\Omega_{Ld}$ report  theoretical values
  calculated   from   Eq.~\eqref{eq:XZ:htheta:def}.   Notice   that   no
  adjustable parameters are introduced  in the calculation. Parameters
  $(\omega,\omega_{0z},\Omega_x,\Omega_z)$  frequencies   in  kHz  and
  $\Phi_{0z}$: in (a) Cs $1.028, 4.42, 3.85, 20.93)$, and $1.63\pi$ as
  in  Fig.~\ref{fig:34};  in (b)  Rb  experiment  $3.0, 77.645,  51.56,
  96.77$, and $\pi/2$; in (c) Cs  $(1.03, 4.66, 4.55, 35.3)$ and 0; in
  (d)  Cs atoms,  (1.028,  4.42,  6.27, 31.79)  and  $0.70\pi$, as  in
  Fig.~\ref{fig:XZTime}c). } 
 \label{fig:XZFreq}
\end{figure}

Our  detection of  the  qubit coherences  leads  to time  dependencies
greatly different from the LZSM transition probabilities.  The greater
contribution    to   the    coherence    oscillation    is   at    the
$\Omega_{Ld}(\tau)$ frequency  and this  frequency describes  also the
probability  oscillations.   As  in  Fig.~\ref{fig:XZTime}(a)  at  low
dressing  values the  qubit spin  performs  a precession  on a  nearly
horizontal  plane at  a  quasi constant  $\Omega_{Ld}$ Larmor  dressed
frequency.  In  those conditions  the transition probability  does not
contain $\Omega_{Ld}(\tau)$ frequency components. At larger $\Omega_z$
values the  Landau-Zener tunneling takes  place in the presence  of a
modification  of   the  precession   axis  leading   to  $\Omega_{Ld}$
components also into $P_+$ probabilities.

In   order  to  characterize   the  qubit  response   at  the
$\Omega_{Ld}(\tau)$   frequency,  we   measure   the  $t_P(j)$   times
corresponding to the $j$-th zero signal values of the polarimeter. For
those times the qubit spin  orientation is orthogonal to the monitored
axis. The time separation  between neighbouring zero values represents
the  period time  for a  single Larmor  precession. From  the measured
times we derive the  experimental dressed Larmor frequency $f_{Ld(j)}$
at the time $t_j=\left[ t_P(j)+t_P(j+1)\right]/2$ given by 
 \begin{equation}
 \label{eq:afL}
f_{Ld}(j)=\frac{2\pi}{t_P(j+1) -t_P(j)}.
\end{equation}
Fig.~\ref{fig:XZFreq} reports  measured $  f_{Ld}$ time  evolutions of
the dressed qubit  frequency for several dressing  parameters. The (a)
plot evidences the $ f_{Ld}$ time periodicity with the dressing period
$T$. Owing to such periodicity all  the $ f_{Ld}$ values may be folded
back into a single $(0,1)$ time period.  Within the temporal analog of
real-space  periodic  Hamiltonian of  Ref.~\citeonline{MartinHalperin_2017},
that  folding  represents  a  reduction   to  a  single  Floquet  zone
corresponding to the Bloch zone  for periodicity in position space for
our  time  periodicity. $  f_{Ld}(t)$  folded  plots are  reported  in
Fig.~\ref{fig:XZFreq}(b) and  (c). Fig.~\ref{fig:XZFreq}(d)  reports a
case where the $ f_{Ld}$ vs  $\tau$ is not fully periodic, as examined
in the following paragraph. The  $f_{Ld}$ measured values are compared
to  the  time  dependent  $\Omega_{Ld}(\tau)$,  showing  a  remarkable
agreement limited by  the time resolution of  the sampling acquisition
rate.  
 \begin{figure}[ht!]
\centering
 \includegraphics [angle=0, width=0.7\textwidth,height=7cm] {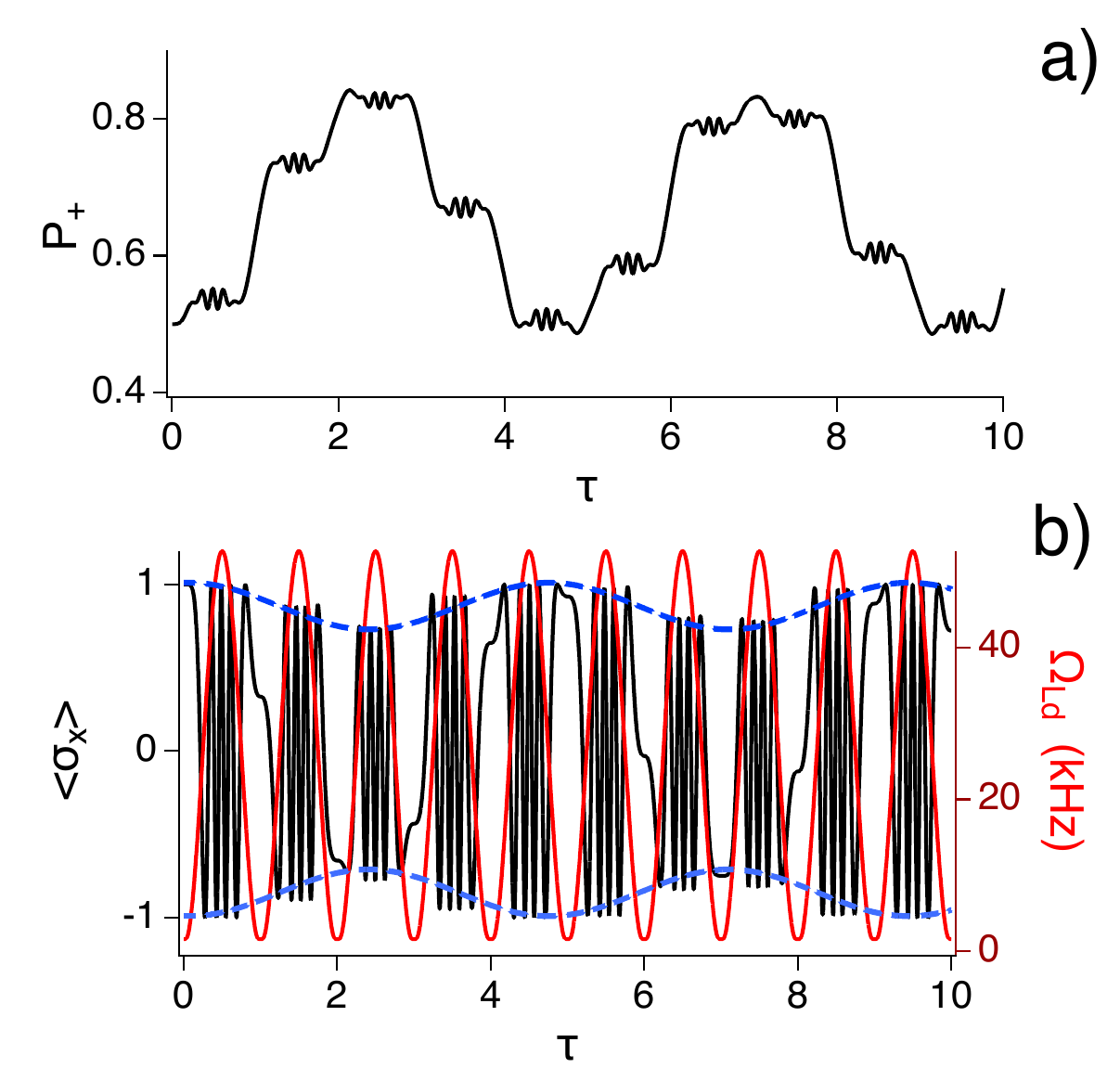}
 \caption{ (Color online) Theoretical simulation of the Rabi-like
   oscillations  with initial  $\langle \sigma_x(0)\rangle=1$.   In a)
   $P_+(\tau)$   occupation    probability   and   in    b)   $\langle
   \sigma_x(\tau)\rangle$. Blue dashed lines are fits of the Rabi-like
   oscillations with $4.61\tau$ oscillation
   period and the $\Omega_{Rl}=0.213$ kHz frequency.
   For the  coherence in  b) the  Rabi-like
   oscillations masked  by the large  amplitude of the  Larmor dressed
   oscillations produce  an amplitude modulation of  the $\Omega_{Ld}$
   oscillations. Parameters  in kHz: $(1., 4.197,  0.254, 4.189)$, and
   $\Phi_{0z}=\pi$.}  
 \label{fig:RabiFreq}
\end{figure}
\indent The  wavefunction   interferences  associated  to   multiple  periodic
Landau-Zener  processes  introduce   additional  periodic  oscillating
structures, for the
$\Omega_z=0$
LZSM  discussed  theoretically  in
Refs.~\citeonline{AshhabNori_07,ShevchenkoNori_10}   and   investigated   in
solid-state  experiments~\cite{ZhouDu_14,BernsOrlando2008,YangLi2021}.
In these works, focused on the occupation probabilities, such periodic
structures   are  denominated   Rabi-like   oscillations  with   their
$\Omega_{Rl}$  frequency  approximated   by  the  $\Omega_z$  dressing
frequency~\cite{AshhabNori_07,NeilingerIlichev_16,LiuLi_21}.       The
Rabi-like  positive and  negative  interferences appear  also for  our
Hamiltonian as in the  simulations of Fig.~\ref{fig:RabiFreq} with the
$P_+(\tau)$   occupation  probability   in   (a)   and  the   $\langle
\sigma_x(\tau)\rangle$  coherence in  (b). Rabi-like  oscillations are
clearly  visible in  the  occupation probability.  The Larmor  dressed
frequency oscillations  produce the  high-frequency modulation  of the
$P_+(\tau)$  envelope. For  the coherence  the Rabi-like  oscillations
produce  an  amplitude  modulation  of the  Larmor  dressed  frequency
oscillations. Their experimental detection  is limited by our sampling
resolution. The  Rabi-like oscillations well  resolved in the  data of
Fig.~\ref{fig:XZTime}(c) produce a periodic  large modification of the
coherence time dependence. Their presence leads to the absence of data
points in the $f_{Ld}$ vs  $\tau$ plot of Fig.~\ref{fig:XZFreq}(d) for
$\tau$ in  the $(0.49,1.06)$ interval, with  $4.13(2)\tau$ oscillation
period, corresponding  to 0.25 kHz.  The experimental results  for the
Rabi-like oscillations  are described by numerical  simulations. Their
dependence on the Hamiltonian parameters appears more complex than the
theoretical ones for the standard LZSM single dressing process.

\section*{Conclusion}

The  time   evolution  of  an   atomic  qubit  interacting   with  two
off-resonant   dressing    fields   in    a   longitudinal/transversal
configuration  is  examined  experimentally. The  qubit  coherence  is
recorded  for  long  interaction   times  with  negligible  relaxation
processes  on   the  evolution   timescales.   The   standard  Floquet
engineering  treatment  predicts  a  time evolution  with  a  constant
dressed  Larmor frequency,  not  matching  these experimental  results
owing to our  dressing parameters.  We develop  a theoretical analysis
relying on  numerical simulations  combined to an  adiabatic treatment
and an ad-hoc dressed-atom perturbation treatment.  {
  An analytical
formula for  the frequency  of the Rabi-like  oscillations will  be an
useful tool.  The complex mathematical treatment for the
$\Omega_z=0$ 
of       Refs.~\citeonline{AshhabNori_07,NeilingerIlichev_16,LiuLi_21}
evidences  a  hard  task to  complete. }  Our  longitudinal/transversal
dressing  configuration   represents  an   extended  version   of  the
multi-passage LZSM  tunneling, with the transverse  dressing modifying
the  tunneling  probability.   An  important feature  of  the  present
experiment is the continuous monitoring of the qubit coherences, which
gives  a  direct  access  to  the phase  wavefunction  and  opens  new
perspectives  in the  LZSM quantum  control.   In terms  of the  Bloch
vector  evolution,   the  transition  probability  monitors   its  $z$
component while the  coherences monitor the evolution  within the $xy$
plane.  The time evolution of the latter shows the presence of several
frequency contributions: the dressing one,  the dressed Larmor one and
the Rabi-like  one. Within the time  dependent transition probability,
the   dressed  Larmor   frequency  components   produce  St\"uckelberg
oscillations,  which  represent  a   single  component  of  the  total
evolution.  On the contrary, those oscillations dominate the coherence
time evolution.   The probability evolution represents  a strong probe
of the Rabi-like oscillations.

Our work  opens a new exploration  direction within the broad  area of
LZSM interferometry. The ability to control the dynamics of the atomic
qubit in  this dual-dressing  regime expands  the toolbox  for quantum
state  manipulation.   Alternative  fast  quantum  logic  gates  using
nonadiabatic     LZSM      transitions     were      introduced     in
Ref.~\citeonline{RyzhovNori_2024}    and    similar    shortcuts    to
adiabaticity                                                        in
Refs.~\citeonline{PetiziolWimberger_2018,PetiziolWimberger_2024}.  Our
experimental investigation evidencing the  rich dynamics associated to
the XZ configuration opens an avenue to quantum control accelerations.






 




\appendix

\section*{Appendices}

\subsection*{Adiabatic perturbation theory}
\label{app:perturbation}

Following the
treatment   of
~\citeonline{DemirplakRice_2005,WeinbergKolodrubetz_2017},         the
adiabatic perturbation theory comes into  play when the Hamiltonian is
slowly changing in  time, i.e., we replace $t  \rightarrow \epsilon t$
where  $\epsilon   \ll  1$   and  $1/\epsilon$  represents   the  slow
characteristic time scale.

Let us  consider a time  dependent Hamiltonian $H(t)$ with  its $G(t)$
diagonalizing unitary operator
\begin{equation}
  \label{eq:def:G}
  G^{\dagger}(t) H(t) G(t) = H_D(t) 
\end{equation}
$H_D$  has a  diagonal form  with $E_i(t)$  instantaneous eigenvalues.
The Schr\"odinger\ equation $(\hbar = 1)$ for the time evolution
operator $U(t)$
\begin{equation}
  \label{eq:def:U}
  i \dot{U}(t) = H(t) U(t) \qquad U(0) = 1
\end{equation}
is rewritten in an adiabatic picture introducing
\begin{equation}
  \label{eq:def:UG}
  U_G(t) = G^{\dagger}(t) U(t) G(0)
\end{equation}
which satisfies
\begin{equation}
  \label{eq:def:UG:din}
  i \dot{U}_G(t) = \big[ H_D(t)  + i \dot{G}^{\dagger}(t) G(t) \big] U_G(t) \qquad U_G(0) = 1
\end{equation}

Introducing $\tau  = \epsilon t$  with $d/dt = \epsilon  d/d\tau$, the
last equation is rewritten as
\begin{equation}
  \label{eq:def:APT}
  i U^{\prime}_G(\tau) = \big[
  \epsilon^{-1} H_D(\tau) + i (G^{\dagger}(\tau))^{\prime} G(\tau)
  \big]
  U_G(\tau), 
\end{equation}
where we have  introduced $f^{\prime} (\tau)=d f(\tau)/d  \tau $.  The
r.h.s.  first  term represents  the  largest  $H_D$ contribution.   We
rewrite the r.h.s. second term as
\begin{equation}
  \label{eq:def:VND}
  i \left[G^{\dagger}(\tau)\right]^{\prime} G(\tau) \equiv V_D(\tau) + V_{ND}(\tau)
\end{equation}
separating the $V_D$ diagonal part related to the Berry phase from the
$V_{ND}$ non-diagonal part.  Separating  the diagonal and not-diagonal
contributions, the $U_G$ evolution operator is given by
\begin{equation}
  \label{eq:def:APT:1}
  i U^{\prime}_G(\tau) = \big[
  \underbrace{\epsilon^{-1} H_D(\tau) + V_D(\tau)}_{\mathrm{diagonal}}
  +
  \underbrace{V_{ND}(\tau)}_{\mathrm{not-diagonal}}
  \big]
  U_G(\tau), 
\end{equation}
\indent By defining the $ Q(\tau)$  operator as the integral  of the diagonal
part
\begin{equation}
  \label{eq:def:Q}
  Q(\tau) \equiv \int_0^{\tau}[ \epsilon^{-1} H_D(\tau) + V_D(\tau)] \de \tau,
\qquad Q(0) = 0,
\end{equation}
we introduce  the $U_{G,I}(\tau)$ adiabatic-interaction time evolution operator  
\begin{equation}
  \label{eq:def:UGI}
  U_G(\tau) = \e^{-i Q(\tau)} U_{G,I}(\tau), \qquad U_{G,I}(0) = 1.
\end{equation}
We obtain 
\begin{equation}
  \label{eq:def:din:UGI}
  \begin{split}
    i (U_{G,I})^{\prime} & = \e^{i Q(\tau)} V_{ND}(\tau) \e^{-i Q(\tau)} U_{G,I}(\tau) \\
    & \equiv \tilde{V}_{ND}(\tau)  U_{G,I}(\tau) 
  \end{split}  
\end{equation}
to be  solved perturbatively.  The potential  $\tilde{V}_{ND}$ induces
transitions between the  eigenstates $|E_i(\tau)\rangle$, neglected in
the $\epsilon\rightarrow 0$ adiabatic limit.

At the lowest perturbation order, equivalent to the adiabatic theorem,
setting     $U_{G,I}(\tau)    \approx     1$     we    neglect     the
$\tilde{V}_{ND}(\tau)$ contribution in Eq.~\eqref{eq:def:UGI} obtaining
\begin{equation}
  \label{eq:def:U0:approx}
U_G(\tau) \approx \e^{-i Q(\tau)}.
\end{equation}
The approximated time-evolution operator results
\begin{equation}
  \label{eq:def:U0}
U(\tau) \approx  G(\tau) U_G(\tau) G^{\dagger}(0) = G(\tau) \e^{-i Q(\tau)} G^{\dagger}(0) 
\end{equation}

\subsection*{XZ rotating Hamiltonian}
\label{app:rotHam}
Let us consider a simplified XZ excitation scheme, denoted as rotating
XZ, where  $\Omega_x=\Omega_z=\Omega$ and the x,z  dressing components
with $\Phi$ phase shift. The Hamiltonian is given by
\begin{equation}
  \label{eq:XZ:H}
  H = \frac{1}{2}
  \left[
    \Omega \sin(\omega t + \Phi) \, \sigma_x +
    (\omega_{0z} + \Omega\cos(\omega t + \Phi)) \,\sigma_z
  \right].
\end{equation}
with    the    $\vec{h}^R$     effective    field,    equivalent    of
Eq.~\eqref{eq:XZ:htheta:def}, given by 
\begin{equation}
  \label{eq:XZ:h:def}
  \vec{h}^R =
  \begin{pmatrix}
    \Omega \sin(\omega t + \Phi) \\
    0\\
    \omega_{0z} + \Omega\cos(\omega t + \Phi)
    \end{pmatrix}.
\end{equation}

Let's define
\begin{equation}
  \label{eq:XZ:def:m}
  m = \frac{\omega_{0z}}{\Omega}
\end{equation}
and introduce the $ h^R$ modulus and the $\theta^R$ vector orientation
angle,
\begin{subequations}
\label{eq:XZ:hmod:def}
\begin{align}
  h^R(t) &\equiv 
      \Omega \sqrt{1 +2 m \cos(\omega t + \Phi) + m^2}\\
  \theta^R(t) &\equiv \arctan\left( \frac{h^R_x}{h^R_z} \right) =
           \arctan
           \left( \frac{\sin(\omega t + \Phi)}{m + \cos(\omega t +\Phi)} \right),
\end{align}
\end{subequations}
this  last  one  equivalent  to  the  $\theta$  orientation  angle  of
Eq.~\eqref{eq:XZ:Eigen:def}.     The     rotating    Hamiltonian    of
Eq.~\eqref{eq:XZ:H} rewritten as
\begin{equation}
  \label{eq:XZ:H:rew}
  H = \frac{h^R}{2} \left[ \cos \theta^R \, \sigma_z + \sin\theta^R \, \sigma_x \right]
\end{equation}
is diagonalized by the following  $G$ operator representing a rotation
around the $y$ axis:
\begin{equation}
  \label{eq:XZ:def:G}
  G = \e^{ -i \theta^R(t) \, \sigma_y/2 }.
\end{equation}
The diagonalized $H_D$ Hamiltonian results 
\begin{equation}
  \label{eq:XZ:HD:def}
  H_D = G^{\dagger}(t) H (t) G(t) = \frac{h^R(t)}{2} \, \sigma_z
\end{equation}
with    the     r.h.s.    of    Eq.~\eqref{eq:def:APT}     given    by
Eq.~\eqref{eq:def:VND} becomes here
\begin{equation}
  \label{eq:XZ:GcG}
  i \dot{G}^{\dagger} G = - \frac{1}{2}\theta^R(\tau)^{\prime}\, \sigma_y = V_{ND}
\end{equation}
consisting only of off-diagonal terms.

Introducing an adimensional time
\begin{equation}
  \label{eq:XZ:tau:def}
  \tau = \frac{\omega t}{2 \pi} \equiv \epsilon t \qquad \omega \approx 0
\end{equation}
the previous Eq.~(\ref{eq:def:APT:1}) becomes
\begin{equation}
  \label{eq:XZ:UG:dyn}
  i (U_G(\tau))^{\prime} = \frac{1}{2}
  \left[
    \frac{2 \pi}{\omega} h^R \,\sigma_z + \theta^R(\tau)^{\prime} \, \sigma_y 
  \right] U_G(\tau). 
\end{equation}
We introduce the $\phi(\tau)$ time accumulated phase angle of the main
text Eq.~\eqref{eq:XZ:Eigen:def}
\begin{eqnarray}
  \label{eq:XZ:PHI}
  \phi(\tau) =&   \frac{2 \pi}{\omega}& \int_0^{\tau} h^R(\tau) \de \tau  \nonumber \\
  = &\frac{2\pi}{\omega}& \Omega\int_0^{\tau}  \sqrt{1 + 2 m \cos(2\pi\tau + \Phi) + m^2} \de \tau, 
\end{eqnarray}
where the  integral can  be expressed through  a second  kind elliptic
integral. Then, we obtain for the $Q$ operator
\begin{equation}
  \label{eq:XZ:Q}
  Q(\tau) =\frac{1}{2} \phi(\tau) \, \sigma_z,
\end{equation}
and for the time evolution operator in the lowest order
\begin{equation}
  \label{eq:XZ:U:fin}
  U(\tau) = \e^{-i \theta(\tau)\,\sigma_y/2}\, \e^{-i\Phi(\tau)\,\sigma_z/2}\,\e^{i \theta(0)\,\sigma_y/2}.
\end{equation}

Using this operator,   the  $\sigma_i$   time evolution is given by
\begin{equation}
\sigma_i(\tau)  =  U^{\dagger}(\tau)  \sigma_i U(\tau).
\end{equation}
Introducing $\theta_0  \equiv \theta(0)$  for the  initial orientation
angle  of the  effective field,  with  the system  at time  $t=\tau=0$
prepared in the $\sigma_x  |\psi\rangle = |\psi\rangle$ eigenstate, we
obtain           for           the           expected           values
$\langle \psi | \sigma_i (\tau) | \psi \rangle \equiv \langle \sigma_i
(\tau) \rangle$
\begin{subequations}
\label{eq:XZ:si:transf:vm}
\begin{eqnarray}
  \langle \sigma_x(\tau) \rangle
  &= &\sin(\theta)\sin(\theta_0)+ \cos(\theta)\cos(\phi)\cos(\theta_0 \nonumber )\\
  \langle \sigma_y(\tau)  \rangle
  &=&   \cos(\theta_0) \sin(\phi) \nonumber\\
  \langle \sigma_z(\tau)  \rangle
  &=& \cos(\theta)\sin(\theta_0)-\sin(\theta)\cos(\phi)\cos(\theta_0),
\end{eqnarray}
\end{subequations}
with the $\phi(\tau)$ phase angle defined by Eq.~\eqref{eq:XZ:PHI} for
the XZ rotating Hamiltonian, and the equivalent definition in the main
text.

\section*{Model for the low-frequency nonadiabatic regime}
\label{appendix:nonad}

In this Appendix, an analytical  approximation for the dynamics in the
low driving frequency, strong  $z$-dressing regime of Fig.~\ref{fig:34} (c) -(d)
is derived,  whose parameters  are reported  here for  convenience (in
kHz):
\begin{align}
  & \omega/2\pi = 1.028, \quad \omega_{0z}/2\pi = 4.42 \\
  & \Omega_x/2\pi = 3.85, \quad \Omega_z/2\pi = 20.93 .
\end{align}
This regime is not well captured neither by an adiabatic approximation
(for which $\omega$ is not small enough) nor by a conventional Floquet
high-frequency   expansion   (for   which  $\omega$   is   not   large
enough).  However, the  fact that  $\Omega_z$ is  substantially larger
than other  parameters can still  be used to construct  a perturbative
expansion, in  a suitable  transformed frame, which  well approximates
the dynamics. 
\begin{figure}[ht!]
  \centering
  \includegraphics[width=0.45\textwidth]{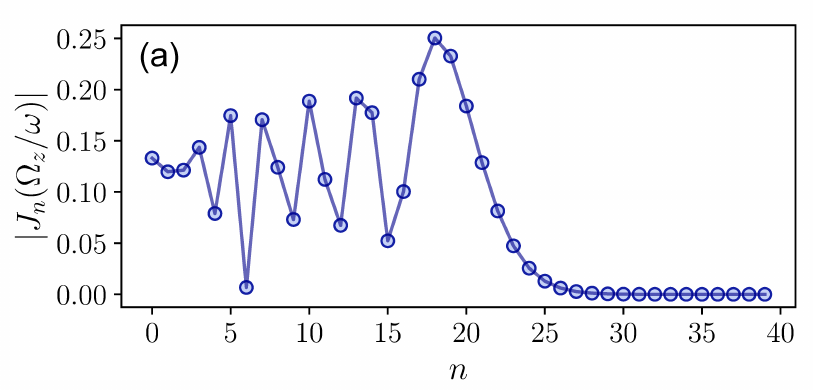}
  \includegraphics[width=0.45\textwidth]{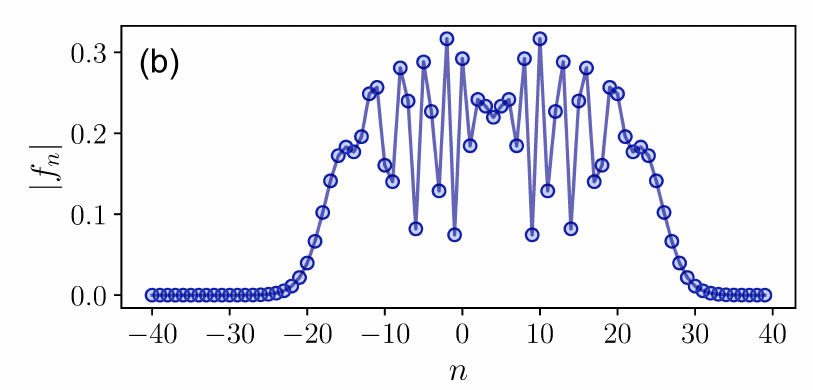}
  \caption{(a)    Absolute    value     of    the    Bessel    functions
  $J_n(\Omega_z/\omega)$ entering the Fourier coefficients of $H''(t)$
  of Eq.~\eqref{eq:Hprimeprime}. (b) Fourier coefficients of $f(t)$ of
  Eq.~\eqref{eq:Hprimeprime2}.} 
\label{fig:C1}
\end{figure}
The transformed frame is defined by  two steps. First, since the qubit
splitting  $\omega_{0z}$ is  close  in value  to  four driving  quanta
$\omega$,  a  rotating frame  oscillating  at  frequency $4\omega$  is
chosen. This is defined by the unitary
\begin{equation} \label{eq:V1transf}
V(t) = e^{-i(4\omega) t \sigma_z /2}
\end{equation}
and the transformed-frame Hamiltonian $H' = V^\dag H V-i V^\dag \partial_t V$ reads as
\begin{equation}
  H'(t) = \frac{\delta}{2}\sigma_z +
  \frac{\Omega_x}{2}\cos(\omega t)\left( e^{i 4 \omega t}\sigma_+
    + e^{-i 4 \omega t}\sigma_+ \right) 
+ \frac{\Omega_z}{2} \cos(\omega t+\Phi_{0z}) \sigma_z,
\end{equation}
with  $\delta =  \omega_{0z}  -  4 \omega$.   The  second  step is  to
integrate  out  (choose a  frame  co-rotating  with) the  $z$-dressing
field, via the transformation
\begin{subequations}
\begin{align} \label{eq:V2transf}
V'(t) & = \exp\left(-i \frac{\Omega_z}{2}\int_0^t dt' \cos(\omega t'+\Phi_{0z})\sigma_z\right) \\
& = \exp\left(-i \frac{\Omega_z}{2\omega} \left[\sin(\omega t+\Phi_{0z}) - \sin(\Phi_{0z})\right]\sigma_z\right) .
\end{align}
\end{subequations}
Note that, although we found  conceptually more clear to introduce the
two steps $V(t)$ and $V'(t)$  separately, they commute with each other
and,  thus,  they can  be  straightforwardly  combined into  a  single
transformation.          The          transformed          Hamiltonian
$H'' = (V')^\dag H' V'-i(V')^\dag \partial_t V'$ is given by
\begin{equation}
  \label{eq:Hprimeprime}
H''(t) = \frac{\delta}{2}\sigma_z
+ \frac{\Omega_x}{2}\cos(\omega t)
\times \Big( e^{i 4\omega t +  i (\Omega_z/2)\left[ \sin(\omega t+\Phi_{0z}) - \sin(\Phi_{0z})\right]}\sigma_+ + \mathrm{H.c.} \Big) .
\end{equation}
By expanding  the exponentials via  Bessel functions~\cite{NIST:DLMF},
$H''$ can finally be rewritten as
\begin{equation} \label{eq:Hprimeprime2}
H''(t) = \frac{\delta}{2}\sigma_z + \frac{\Omega_x}{4} f(t) \sigma_+ +\frac{\Omega_x}{4} f^*(t) \sigma_-,
\end{equation}
where $f(t)=\sum_{n=-\infty}^{\infty} f_n \exp(in\omega t)$ has Fourier components
\begin{equation}
  \label{eq:fn}
f_n = \Big[J_{n-5}\left(\frac{\Omega_z}{\omega}\right) e^{-i 5 \Phi_{0z} } + J_{n-3}\left(\frac{\Omega_z}{\omega}\right) e^{-i 3 \Phi_{0z} }\Big]
\times e^{i n\Phi_{0z}} e^{-i\Omega_z\sin(\Phi_{0z})/\omega} .
\end{equation}
The      Fourier     components      of     $H''$,      defined     by
$H''=\sum_{n=-\infty}^{+\infty}  H_n \exp(in\omega  t)$,  can thus  be
identified as
\begin{equation} \label{eq:Hn}
H_n = \frac{\delta}{2}\sigma_z\delta_{n,0}+ \frac{\Omega_x}{4} \left[ f_n \sigma_+ + f_{-n}^* \sigma_- \right],
\end{equation}
with $f_n$ given  in Eq.~\eqref{eq:fn}.  In the  transformed frame, we
are  still  dealing  with  a time-periodic  Hamiltonian  at  frequency
$\omega$  [Eq.~\eqref{eq:Hprimeprime}].  However,   since  the  Bessel
functions have  small value for large  argument $\Omega_z/\omega$, the
advantage of the  transformed frame is that the  coefficients of $H_n$
have now small magnitude as compared to $\omega$ (differently from the
lab-frame   Hamiltonian).  This   is  illustrated   quantitatively  in
Fig.~\ref{fig:C1},  where the  values  of $|J_n(\Omega_z/\omega)|$  and
$|f_n|$  are  reported  for  the   parameter  values  of  interest  of
Fig.~\ref{fig:34}.     Provided     $\Omega_x/\Omega_z\ll    1$     and
$\delta/\Omega_z\ll 1$,  this allows  one to construct  a perturbative
expansion with standard methods.  In particular, the dynamics $U''(t)$
generated  by  $H''(t)$  can  then be  estimated  by  using  Floquet's
theorem~\cite{GoldmanDalibard_2014,             BukovPolkovnikov_2015,
  Eckardt_2017},
\begin{equation} \label{eq:uppfloquet}
U''(t) = e^{-i K(t)} e^{-i H_{\rm eff}(t-t_0)}e^{i K(t_0)},
\end{equation}
and   a  Floquet   inverse-frequency  expansion   for  the   effective
Hamiltonian      $H_{\rm       eff}$      and       kick      operator
$K(t)$~\cite{GoldmanDalibard_2014,              BukovPolkovnikov_2015,
  Eckardt_2017}. To  second-order, this  is computed from  the Fourier
components  $H_n$ of  Eqs.~\eqref{eq:Hn}  and \eqref{eq:fn}  according
to~\cite{Eckardt_2017}
\begin{align}
  & H_{\rm eff} \simeq H_0 + \frac{1}{\omega}\sum_{n=1}^\infty \frac{1}{n}[H_n, H_{-n}] \\
  &K(t) \simeq -\frac{i}{\omega}\sum_{n\ne 0}^\infty \frac{H_n}{n}e^{in\omega t} +\frac{i}{\omega^2}\sum_{n\ne0} \left(\frac{[H_0, H_n]}{n^2} e^{in\omega t}+ \sum_{m\ne0,n} \frac{[H_{-m}, H_n]}{2n(n-m)} e^{i(n-m)\omega t} \right).
\end{align}
Note, once more, that the  small parameter justifying the expansion is
not  the  inverse  frequency  $\omega^{-1}$, but  rather  (the  matrix
elements of) $H_n$. The lab-frame evolution $U(t)$ is finally obtained
as
\begin{equation} \label{eq:labframe}
U(t) = V(t) V'(t) U''(t)
\end{equation}
from  Eqs.~\eqref{eq:V1transf}   and~\eqref{eq:V2transf}.  Since  both
$V'(t)$ and $V''(t)$ are $\omega$-periodic,  they can be absorbed into
the  micromotion  operator,  such  that  Eqs.~\eqref{eq:labframe}  and
\eqref{eq:uppfloquet}   straightforwardly    define   a   Floquet-type
decomposition also  in the  lab frame.  Given  that the  parameters of
Fig.~\ref{fig:34}  satisfy the  condition $\Omega_x/\Omega_z\ll  1$ and
$\delta/\Omega_z\ll   1$,   which   underlie  the   expansion,   quite
approximately, it  is to be  expected that a  satisfactory description
requires beyond-first-order  terms in  the expansion.  We find  that a
second-order  truncation  already  reproduces fairly  well  the  exact
dynamics.


\bibliography{biblioDressingINO}



\section*{Acknowledgements}

\noindent All authors thank D.~Petrucci and A.~Barbini for technical assistance.

\noindent S.~W.     acknowledges    financial     support    by:  Q-DYNAMO (EU
HORIZON-MSCA-2022-SE-01) with project No.~101131418. 

\noindent A.~Fr. acknowledges the
Italian MUR through the PNRR project ``EuPRAXIA Advanced Photon Sources
(EuAPS)'' contract EuAPS IR0000030 CUP I93C21000160006 for the support. 

\noindent A.~Fi. acknowledges financial support from  the PNRR MUR project PE0000023-NQSTI financed by the European Union - Next Generation EU

\noindent F.~P. acknowledges funding from the Deutsche Forschungsgemeinschaft (DFG, German Research Foundation) through the Emmy Noether Programme -- project number 555842149.

\section*{Author contributions statement}
A.~Fr., A.~Fi.,  C.~G., C.~M.  and V.~B.  performed the  experiments: E.~A.,
V.~B. and A.~Fi.  conceived the  probe tools. A.~Fr., V.~B.  and A.~Fi.  analysed
the  results.   E.~A.,  M.~A., G.~B., S.~W.    and  F.~P.   performed  the  theoretical
analysis.   All  the authors  prepared  the  manuscript.  All  authors
reviewed the manuscript.








\end{document}